\documentclass[sigconf,nonacm]{acmart}

\usepackage{soul}

\AtBeginDocument{%
  }

\setcopyright{none}
\copyrightyear{2026}
\acmYear{2026}

\acmConference[Agentic Software Engineering (SE 3.0) Workshop @ KDD'26]{Agentic Software Engineering (SE 3.0) Workshop, co-located with the 32nd ACM SIGKDD Conference on Knowledge Discovery and Data Mining}{August 9--13, 2026}{Jeju, Korea}

\renewcommand\footnotetextcopyrightpermission[1]{}
\begin{document}

\title{Position: Coding Benchmarks Are Misaligned with Agentic Software Engineering}

\author{Maria I. Gorinova}
\affiliation{%
  \institution{Tessl}
  \city{London}
  \country{UK}
}
\email{maria@tessl.io}

\author{Macey Baker}
\affiliation{%
  \institution{Tessl}
  \city{London}
  \country{UK}
}
\email{macey@tessl.io}

\author{Amy Heineike}
\affiliation{%
  \institution{Tessl}
  \city{London}
  \country{UK}
}
\email{amy@tessl.io}

\author{Maksim Shaposhnikov}
\affiliation{%
  \institution{Tessl}
  \city{London}
  \country{UK}
}
\email{max@tessl.io}

\author{Rob Willoughby}
\affiliation{%
  \institution{Tessl}
  \city{London}
  \country{UK}
}
\email{robw@tessl.io}

\author{Dru Knox}
\affiliation{%
  \institution{Tessl}
  \city{London}
  \country{UK}
}
\email{dru@tessl.io}

\renewcommand{\shortauthors}{Gorinova et al.}

\begin{abstract}
Coding agents have become a major mode of software engineering, but the
benchmarks we use to compare them were designed in a pre-agent era:
they collapse model, harness, and environment into a single end-to-end
score, typically computed against one reference solution, with no
component-level signal for iteration. \textbf{We argue that current coding benchmarks are misaligned with
agentic software engineering.} A coding agent in practice is not a
model: it is a \emph{system harness} --- a composite of models, harnesses,
contexts, environments, and feedback signals, any one of which can move
the benchmark score by margins comparable to those between adjacent
model generations. We discuss three symptoms: (i) benchmark scores
conflate the model with the rest of the harness; (ii) grading against a
single reference solution penalises equally valid alternatives; and
(iii) the absence of signal at the level of individual harness components
makes the end-to-end system score difficult to iterate on.
\end{abstract}

\maketitle

\section{Introduction}

Coding agents \cite{claude-code, codex, cursor-agent, yang2024sweagent, wang2025openhands} are now a major mode of software engineering. They open
and merge pull requests, write internal libraries, and increasingly
take on multi-day engineering work under human
supervision. They are composite systems, consisting of a large language model (LLM) in a tool-use loop, scaffolding, environment, and context. Each of these components can shift the
end-to-end benchmark score by margins comparable to those between
adjacent model generations~\cite{morphllm2025swepro,
ai21scalingeval2025}. 

But the benchmarks we use to
compare them were designed for an earlier object of study: the ability of an LLM to generate working code in one go. SWE-Bench~\cite{jimenez2024swebench},
HumanEval~\cite{chen2021humaneval}, MBPP~\cite{austin2021mbpp},
LiveCodeBench~\cite{jain2024livecodebench}, and
BigCodeBench~\cite{zhuo2024bigcodebench} all share the same structure:
a single model, a single harness, and a single environment together
produce a single number --- an end-to-end system score with no signal at the
level of individual components --- which is often compared against a single reference solution.

The choice of benchmark is not neutral: it implicitly shapes how methods are judged and which research directions get pursued, even if the benchmark only partially captures the unit of interest~\cite{dehghani2021benchmarklottery}.

\textbf{We take the position that current coding benchmarks
are misaligned with agentic software engineering: they grade only a small part of what we build, against constructs we do not want.} Closing the gap requires benchmarks designed around the structure of agentic systems,
rather than around individual reference solutions. Such benchmarks
would treat the agent as the composite system it is, expose signal at
the level of individual components, and ground correctness in
independent behavioural specifications rather than in any single
reference solution.
The hardest open problem inside this programme is \emph{operationalisation}: specifying what we want the system to do in terms that can be measured, without encoding how the agent should attempt it.

\section{The System Harnesses}
\label{sec:harness}

\begin{figure}[t]
  \centering
  \includegraphics[width=\linewidth]{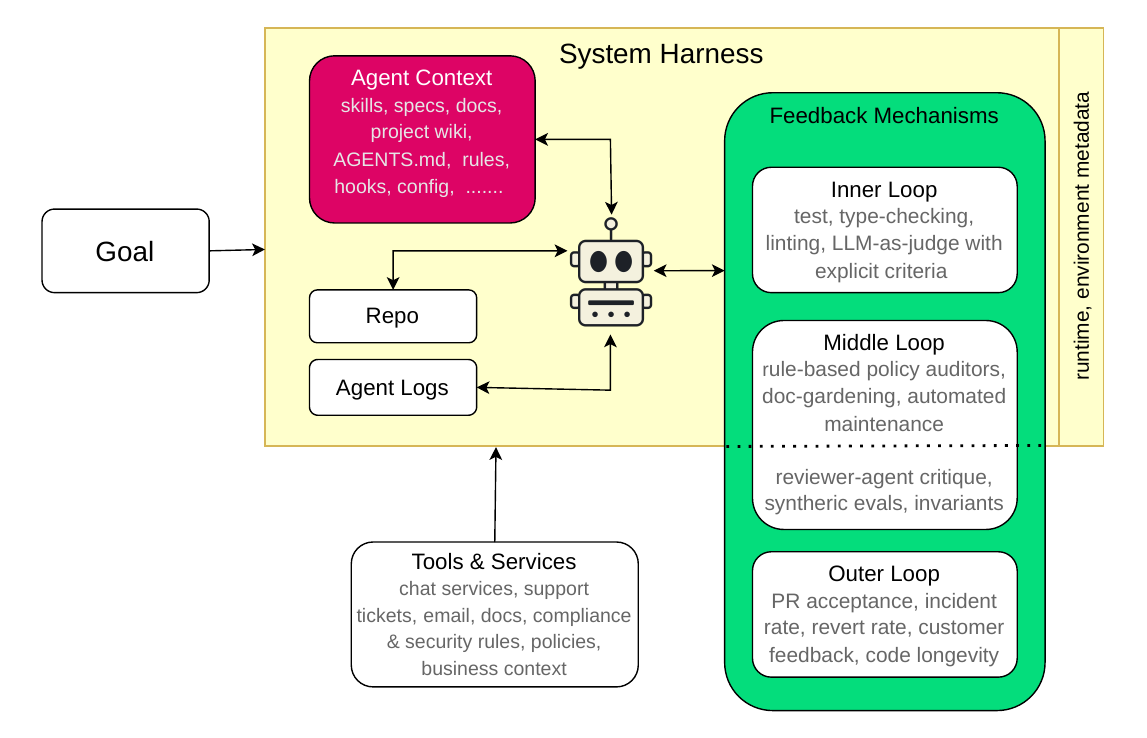}
  \caption{The system harness around a coding agent.
  \textbf{Yellow}: components the harness modifies or produces.
  \textbf{Outside}: what it reads but does not control. Feedback
  signals split into \mbox{inner-,} middle-, and outer-loop tiers; each tier
  also splits into agent-controlled (the harness can rewrite the
  check) vs.\ external (PR comments, production outcomes).}
  \label{fig:harness}
\end{figure}

A coding agent is not a model: it is part of a
\emph{system harness}, an orchestration layer around one or more
LLMs that manages tasks, environments, and feedback over
time. We distinguish two levels of this orchestration. The \emph{agent
harness} is a language model interacting with tools,
working towards a single task, with some system prompt and context to
draw on. Most artefacts described as ``coding agents'' are agent
harnesses in this sense  ---  Claude Code \cite{claude-code}, Codex \cite{codex}, Cursor Agent \cite{cursor-agent}, SWE-Agent \cite{yang2024sweagent}, OpenHands \cite{wang2025openhands},
and many others. The \emph{system harness} is the outer orchestration: it
transforms higher-level goals into concrete tasks, dispatches each
to one or more agent harnesses, manages the environment they act on,
and routes their outputs through feedback that approximates whether
the work is acceptable. Practical agentic coding at scale operates
at the level of the system harness~\cite{openai2026harness, anthropic2025harness}; current coding benchmarks operate
at the level of the agent harness. Recent examples include
Symphony~\cite{symphony2026} and GasCity~\cite{gascity}. We also built and open-sourced NS2,\footnote{\url{https://github.com/drufball/ns2}} an issue-driven system harness running a four-tool agent loop under a stack of deterministic and agent-arbitrated checks. In NS2, GitHub issues are the coordination primitive: a product-manager agent decomposes a feature issue into vertical-slice child issues; software-engineering agents implement those issues in isolated sessions; reviewer, test-quality, smoke-testing, and PR-building agents consume issue and pull-request events to decide whether work should move forward, be revised, or be described for review. Building and operating NS2 surfaced many of the misalignments this paper articulates, and we draw on it as a concrete reference throughout. Many harnesses, including NS2, treat the issue tracker
as the durable state machine for work and spawn per-task agent sessions.

The system harness has five recurring components
(Figure~\ref{fig:harness}): (i) \emph{tasks}, units of work derived
from higher-level goals; (ii) one
or more \emph{agent harnesses}, configurable executors composed of
model, prompt, tools, and loop, which the system harness may tune or
treat as black boxes; (iii) the \emph{environment} the repository and runtime under change, together
with integrated external services (issue tracker, CI,
deployment surface); (iv) \emph{context}, a curated projection of the
environment and of harness-authored material --- skills, plugins,
hooks, specs --- loaded into a particular invocation;
and (v) \emph{feedback signals}: anything the harness reads to refine
a solution or to refine itself, including tests, types, linters,
formal verification, LLM-as-judge rubrics, PR comments, reviewer
critique, production incidents, and longer-horizon business signals.
Within feedback, \emph{verifiers} are the strict subset that return a
pass/fail verdict suitable for blocking --- tests, type-checks, linters,
and binary rubrics; the broader feedback category includes qualitative
review and outcome signals that inform rather than gate. 

We categorise feedback into three tiers by scope, latency and trust.
\emph{Inner-loop} signals (seconds to minutes: tests, types, lint,
compile) are fast and cheap but narrow. They operate inside a single
change attempt, usually at commit or pull-request scope, and give the
agent immediate steer on whether the current edit is executable,
well-typed, policy-compliant, or worth further iteration.
\emph{Middle-loop} signals (minutes to hours: reviewer requests,
simulation, maintenance agents, score rubrics) aggregate over a wider
slice of work. A \emph{maintenance agent} is a scheduled or on-demand
agent that inspects a project, its PRs, or its agent traces, then files
or implements recurring quality improvements to the software project. Middle-loop signals
capture properties that are too broad or too
judgement-dependent for a unit test: recurring review criticism, drift
from project conventions, duplicated abstractions, agent inefficiency
visible in logs, or before/after health metrics for a maintenance run.
\emph{Outer-loop} signals (days to weeks: PR acceptance, revert rate,
incident reports, customer feedback) are closest to ground truth but
delayed and confounded; they measure whether shipped work survived
human review, production, and use, often without clean attribution to
one prompt, task, or change. A productive system harness uses all
three: inner signals to refine a solution in-loop, middle signals to surface recurring issues and quality drift, and outer signals to calibrate
which inner and middle proxies are worth trusting. Middle-loop
signals are therefore useful when they predict or improve outer-loop
outcomes such as fewer reverts, fewer review rounds, or a lower
human-intervention rate.
A second, orthogonal axis distinguishes signals the harness can modify
(e.g. tests) from signals it
cannot (human PR comments, business outcomes). All signals can take part in the harness's \emph{self-improvement} loop, in which
accumulated logs and recurring failures feed back into the harness's
own components~\cite{lee2026metaharness, openai2026harness}. NS2
illustrates the pattern: max-pedantic lint, a coverage threshold, and
dependency-graph unit tests act as inner-loop strict verifiers;
mutation testing, LCOM cohesion checks, and daily agentic architecture
and test-quality reviews are middle-loop feedback; friction reports
from a smoke-testing agent, post-merge revert signals, and production
incidents are outer-loop feedback. The agent writes and maintains the
lint rules and rubrics that constrain it.

\section{Related Work}
\label{sec:related}

We read existing coding-agent evaluation work through the lens of the system harness
\S\ref{sec:harness}.

\textbf{Coding Benchmarks.} Coding benchmarks fall into two
families. The first scores models on short, self-contained problems
with hidden tests: HumanEval~\cite{chen2021humaneval},
MBPP~\cite{austin2021mbpp}, LiveCodeBench~\cite{jain2024livecodebench}
(with contamination control), and
BigCodeBench~\cite{zhuo2024bigcodebench}. These were designed when
the artefact under test was a model, and they correctly target the
model component of the agent harness. They are not designed to
discriminate between system harnesses. The second
family grades agents on patches to real repositories.
SWE-Bench~\cite{jimenez2024swebench} requires an agent's patch to
make a held-out \texttt{FAIL\_TO\_PASS} set pass while keeping a
\texttt{PASS\_TO\_PASS} set passing; both test sets are derived from
the original pull request, a construction we return to in
\S\ref{sec:human-anchoring}. The benchmark has since iterated to
address distinct shortcomings:
Verified~\cite{openai2024swebenchverified} curates 500
human-validated tasks, Multimodal~\cite{yang2025swebenchmm} adds
visual domains, and Pro~\cite{swebenchpro2025} broadens the task
horizon and language coverage with human-validated tasks (and is now
recommended by OpenAI in place of Verified~\cite{openai2026swebenchretire}).
SWE-rebench~\cite{badertdinov2025swerebench} trades human validation
for scale, generating 32k+ tasks across 20 languages. Beyond the issue-fix shape,
Terminal-Bench~\cite{merrill2026terminalbench} broadens evaluation
to terminal tasks (typically 1--20 minutes) and has become a
de-facto frontier reporting standard;
Frontier-SWE~\cite{frontierswe2026} targets ultra-long-horizon
performance and ML-research challenges. Adjacent suites include
SWE-Lancer~\cite{openai2025swelancer},
RE-Bench~\cite{wijk2024rebench}, MLE-bench~\cite{chan2024mlebench},
$\tau$-bench~\cite{yao2024taubench},
AgentBench~\cite{liu2024agentbench}, and the
Aider polyglot~\cite{aiderpolyglot}. These benchmarks vary the task
domain, time horizon, and verifier shape, but share the same set-up:
the agent is paired with a fixed environment and verifier,
and a single end-to-end pass rate is reported. In the language of
\S\ref{sec:harness}, the rest of the system harness is folded into
the protocol rather than treated as part of the artefact under test.

\textbf{Validity and the Benchmark--deployment Gap.} Recent papers
expose validity problems in the SWE-Bench-style set-up:
SWE-Bench+~\cite{aleithan2024swebenchplus} documents solution leakage
in issue text and weak-test passes;
\citet{swebenchillusion2025} report file-localisation behaviour
consistent with memorisation; \citet{wang2025solvedissues} use
differential testing to show 7.8\% of resolved patches fail
developer-written tests and 29.6\% diverge from the gold patch's
behaviour; and
\citet{many-swe-bench-passing-prs-would-not-be-merged-into-main}
find that many resolved patches would not be merged under ordinary
maintainer review. \citet{li2025aidev,li2026aidev} measure the gap at the other
end: across 456k agent-authored PRs in 61k repositories, real-world
acceptance rates are 35--64\% --- well below the $>$70\% headline
figures on Verified. Their AIDev dataset is useful beyond the gap
it documents. Because it is drawn from live repositories rather than
a curated benchmark, it can report acceptance, review turnaround, and
code complexity in place of a single pass rate, which is closer to
the measurement we argue for. A separate line of work has begun to treat the harness itself
as the object of measurement. \citet{fan2025sweeffi} report that the
same scaffold varies 3--7$\times$ in token-budget effectiveness
across base LLMs and conclude that effectiveness is a property of
the scaffold--model integration, not of either component alone.
SkillsBench~\cite{skillsbench2026} measures the lift attributable to
agent skills, and Meta-Harness~\cite{lee2026metaharness} treats the
harness as an object of optimisation, searching the harness-code
space with an outer proposer. Our position is consistent with this
trajectory: if the harness is the artefact, the harness is what
should be measured.

\textbf{Agentic Software Engineering.} The framing of an
agentic software engineering (SE) discipline is taking shape in parallel.
\citet{hassan2025agentic} call for an ``SE 3.0'' research roadmap
around structured human--agent collaboration, with \emph{merge-readiness
packs} replacing test-pass-as-success --- ``passing tests alone is no
longer enough''. Our argument is the measurement counterpart: if the
artefact is a composite system, the benchmark must score the
composite system. We draw on work that treats evaluation as a
measurement problem. \citet{wallach2025position} argue that
evaluating GenAI is a social-science measurement challenge,
distinguishing the \emph{construct} (e.g.\ ``solves the bug'') from
the \emph{operationalisation} (\emph{how} it is measured);
\citet{jacobs2021measurement} formalise the chain of validity a measurement must
satisfy to be informative about its construct. We use this language
directly in \S\ref{sec:symptoms}: single-reference anchoring
(\S\ref{sec:human-anchoring}) is a \emph{content-validity} claim;
bundle conflation (\S\ref{sec:conflation}) is a
\emph{discriminant-validity} claim --- a benchmark that cannot separate
model from harness is measuring something, but not the thing it
labels.

\section{Three Symptoms of the Misalignment}
\label{sec:symptoms}

We discuss three symptoms of the misalignment between current
benchmarks and agentic coding systems.

\subsection{Conflating the Model with the Harness}
\label{sec:conflation}

\begin{table}[t]
  \caption{Entries from the TerminalBench \cite{merrill2026terminalbench} leaderboard showing 
  success rates for Claude Opus 4.6
  across agent harnesses on a fixed task distribution. Within a
  fixed task distribution, success rates can vary by 20 percentage
  points or more --- a range comparable to differences between model generations.}
  \label{tab:conflating}
  \small
  \setlength{\tabcolsep}{6pt}
  \renewcommand{\arraystretch}{1.1}
  \begin{tabular}{@{}r l l l r@{}}
    \toprule
    Rank & Agent & Model & Agent Org & Accuracy (\%) \\
    \midrule
    4  & ForgeCode     & Opus 4.6 & ForgeCode      & $79.8 \pm 1.6$ \\
    8  & Capy          & Opus 4.6 & Capy           & $75.3 \pm 2.4$ \\
    11 & Terminus-KIRA & Opus 4.6 & KRAFTON AI     & $74.7 \pm 2.6$ \\
    14 & TongAgents    & Opus 4.6 & Bigai          & $71.9 \pm 2.7$ \\
    17 & Droid         & Opus 4.6 & Factory        & $69.9 \pm 2.5$ \\
    20 & Crux          & Opus 4.6 & Roam           & $66.9$ (N/A)   \\
    22 & Mux           & Opus 4.6 & Coder          & $66.5 \pm 2.5$ \\
    28 & Terminus 2    & Opus 4.6 & Terminal-Bench & $62.9 \pm 2.7$ \\
    40 & Claude Code   & Opus 4.6 & Anthropic      & $58.0 \pm 2.9$ \\
    \bottomrule
  \end{tabular}
\end{table}

This conflation is not new: \citet{dehghani2021benchmarklottery} made
a closely related point about non-agent benchmarks, and the SWE-Bench
community has rediscovered it incrementally since. Our position is
that it has not been sufficiently acted on; the cost of inaction has grown because
the model is a small part of what gets used in practice, and the remedy
likely needs structural change at the benchmark level, alongside individual care.

Table~\ref{tab:conflating} shows success
rates for a single fixed model (Claude Opus 4.6) across several agent
harnesses on Terminal-Bench, a benchmark of terminal-based tasks in
containerised environments. Each task gives the agent an English
instruction, a sandboxed environment, and hidden tests; solving it
requires ordinary terminal work such as inspecting files, running
commands, debugging failures, editing code or configuration, and
recovering from intermediate mistakes~\cite{merrill2026terminalbench}.
On this fixed task distribution, differences of 20
percentage points or more appear across harnesses. Practitioner
reports document 4--10 point swings for Claude Opus 4.5 between
standardised and custom scaffolds on SWE-Bench
Verified~\cite{morphllm2025swepro}, and the OpenHands harness reaches
77.6\% with comparable models that score several points lower under
the standardised mini-SWE-Agent harness~\cite{wang2025openhands}.
The effect is not just additive: \citet{fan2025sweeffi} report that
effectiveness ``is not an inherent property of the scaffold'' --- it
emerges from how the scaffold integrates with the base model, with
model swaps moving resolve rate 2--3$\times$ at fixed scaffold.
Across more than 200{,}000 SWE-Bench runs,
\citet{ai21scalingeval2025} find that orchestration choices,
container allocations, and evaluation seeds materially move the pass
rate at fixed model and fixed harness; Anthropic
report similar infrastructure-level noise inside Anthropic's own
evaluation pipeline \cite{harnessvariance2026}. 

The model is fixed across the rows, so the spread cannot be explained
as a difference in model capability. It is showing that the agent
harness --- prompt, tool interface, action loop, environment handling,
retry behaviour, and conventions for using the terminal --- is part of
the measured object. A model may also have been trained or tuned under
particular tool-use conventions, so a harness can be well or poorly
matched to the model even before any task-specific reasoning begins.
Another under-specified variable is inference effort: some model APIs
let the caller directly change the amount of inference compute used to
produce an answer, trading thoroughness against speed and cost,
including across tool calls. Changing this setting can therefore move
the quality of the output even when the model name and agent code are
nominally unchanged.

But recent work often publishes single-harness, single-number comparisons regardless.
The result is attributed to the model, while it is a property
of the agent harness and environment as a whole; the model is one
component among several. A leaderboard entry of the form ``Model
$M$, 65\% on SWE-Bench Verified'' is uninformative about whether $M$
would resolve a given test under different scaffolding or environment; comparing two such numbers is comparing two systems, not two
models. End-to-end numbers are informative, but we
claim they are \emph{under-specified}, and that the unit being measured is not the unit being used in practice.

\subsubsection*{\textbf{Suggested remedy}}

The fix is structural rather than
methodological. Leaderboard maintainers and benchmark stewards should
require  relevant metadata at submission: what model, agent harness version,
environment hash, and dataset version were used. Additionally, submissions should include at least one ablation across a
non-model axis against a fixed baseline.

\subsection{Anchoring on a Single Reference Solution}
\label{sec:human-anchoring}

Many coding benchmarks grade a solution by its closeness
to a single reference. SWE-Bench is the canonical instance for
agentic work~\cite{jimenez2024swebench}: the \verb|FAIL_TO_PASS|
and \verb|PASS_TO_PASS| test sets are derived from the test files
modified in the original pull request, encoding a particular
decomposition  ---  which functions exist, their signatures, etc. An agent that resolves a flaky test by
restating the API at a different level of abstraction is judged
not on whether the bug is fixed, but on whether the reference tests
still hold. In measurement terms, the patch is a proxy for the
construct~\cite{jacobs2021measurement, wallach2025position}.

Such grading is fair only when we specify tasks tightly enough that the agent has no real choice but to
make the same implementation decisions as the reference solution.
In practice, work is rarely this tight, and it is rarely just
bug-fixing: developers ask agents to \emph{define} the API, upgrade
dependencies, evolve abstractions, or choose between architectural
shapes. Practitioners consistently identify \emph{spec quality}
rather than model capability as the primary
bottleneck~\cite{strongdm2025factory, scanlan2026intercom}.
SWE-Bench instances, by contrast, are selected for tractability:
each comes with a clearly filed issue and a cleanly merged patch.
The benchmark is therefore doubly anchored: the output is graded
against the reference patch, and the input is pre-selected to the
standard of a well-formed issue.

This construction also embeds known weaknesses.
\citet{aleithan2024swebenchplus} report 32.67\% solution leakage in
issue text and 31.08\% passes under insufficient tests;
\citet{wang2025solvedissues} show via differential testing that
7.8\% of resolved patches fail developer-written tests and 29.6\%
diverge from the gold patch's runtime behaviour;
\citet{swebenchillusion2025} document file-localisation consistent
with memorisation. 

A deeper issue is that single-reference grading mistakes both the
construct and the grain. The reference encodes one solution among
many: we want agents that can refactor, restructure, and pick among
reasonable alternatives. And, in narrow domains such as compiler
optimisation or kernel autotuning, find shapes no reference patch
encodes. The shortcomings of single-reference grading are well
established in machine learning more
broadly~\cite{sutton2018rlbook}. 

The hidden unit tests also grade only local behaviour. They cannot
see what distinguishes good code from working code: choice of
abstractions, architectural fit, system design.
An agent can pass every test while degrading the codebase in ways a
reviewer would reject on sight. The methodological move is to grade
not on closeness to a particular solution but on a broader definition of functional correctness and on design-level quality (code is reused rather than duplicated, new
abstractions follow project conventions, the dependency graph stays
sound). These are \emph{invariants} --- conditions that should hold
across many candidate solutions, and across many PRs in the same
codebase. Recent practitioner work moves in this direction:
skill-adherence evaluations~\cite{evaluating-skills} grade against
separately-authored policies; abstraction-adherence
checks~\cite{abstraction-adherence} verify structural properties
without prescribing an implementation;
ProgramBench~\cite{yang2026programbench} grades via agent-generated
behavioural tests rather than source-code comparison. All three
decouple the verifier from any particular candidate solution. The
remaining work is articulation: specifying invariants as rubrics
that can be graded reliably, and choosing tasks for which those
invariants apply. We claim this is the central open problem in
agentic-coding evaluation --- specifying \emph{what} we want the
system to do in terms that do not encode \emph{how}.

\subsubsection*{\textbf{Suggested remedy}}

Replace single-reference-derived test sets
with multi-shape behavioural verifiers --- property tests, reference
oracles, or differential tests against alternative implementations.
Where a single gold patch is retained, declare which behaviours are
required and which are incidental to the reference implementation.

\subsection{The Absence of Component-Level Signal}
\label{sec:components}

End-to-end agent runs on a single benchmark task can take hours \cite{frontierswe2026}, yet each task yields only a small amount of signal.
As we saw in \S\ref{sec:harness}, a modern agentic system --- e.g. the one described by \citet{openai2026harness}, or NS2 with its stack of linters, dependency-graph unit tests, mutation testing, agentic reviewers, and a smoke-testing agent --- has many components, each affecting the overall result. If one component is failing, the evaluation of an end-to-end task will capture the failure, but we will not necessarily be able to tell which component is faulty.
To determine how to improve the overall harness,
we might need to resort to running ablation experiments, thus making the improvement loop even more time-consuming.

Practitioners building toward
autonomous systems describe a continuous improvement cycle in which failures must be diagnosed and fixed at the component level:
context, tooling, verifier, or task decomposition~\cite{openai2026harness,
scanlan2026intercom}. An end-to-end score shows that something
failed; it does not say what to fix. Without component-level signal,
this cycle degrades to intuition-guided ablation.
If the harness is a composition of components, we should aim to evaluate components separately. This is the same logic that splits software
testing into unit and integration tests~\cite{beck2002tdd}: an
integration test reflects deployment more faithfully but does not
say \emph{which} component broke.

Recent evaluation work moves in this direction on the task side:
\citet{ribeiro2020checklist} decompose tasks into capabilities with
unit-style assertions; \citet{dehghani2021benchmarklottery} show that
aggregated scores conceal which tasks drive the ranking. But the
system under test is still a black box. The same decomposition
should apply to the system itself: each component evaluated in
isolation, as well as in composition.

Component-level evaluation for agentic systems is sparse, especially
for coding. LLM-only benchmarks (e.g.\ one-shot code generation)
effectively evaluate the model component, and a few techniques
evaluate skills as a stand-alone
context~\cite{skillsbench2026, evaluating-skills}. Recent work in
adjacent literature begins to target individual components:
PEEK~\cite{gu2026peek} scores agents on long-context aggregation and
in-context learning rather than end-to-end completion, treating
orientation knowledge as an evaluable artefact;
DecisionBench~\cite{gao2026decisionbench} evaluates how well an agent
delegates sub-tasks across a pool of models. Neither targets coding
directly, but both illustrate the shape: per-component verifiers
that hold the rest of the harness constant. Some harness components
are themselves evaluation targets for others: in NS2, mutation testing
evaluates the quality of a unit-test suite, and an agentic
linter-quality review evaluates the lint configuration.
The harness components form a stack of verifiers-of-verifiers;
reporting only the end-to-end pass rate flattens this stack.

\subsubsection*{\textbf{Suggested remedy}}

Treat the components of Fig.~\ref{fig:harness} as evaluation targets in their own right, answering questions such as ``How effective is the context in aiding the agent?'', ``How well does the agent follow agreed invariants?'', and ``How effective is the agent in converting policy to deterministic verifiers?''. For maintenance agents, report before/after deltas on scoped health metrics --- complexity, duplication, dead code, dependency cycles, or churn hotspots --- rather than only whether the final PR merged.

\section{Alternative Views}
\label{sec:alt}

\textbf{``End-to-end scores reflect real
usage.''} We agree. We are arguing against using \emph{only} end-to-end metrics and
against treating the resulting score as a property of the model
rather than of the harness. In
software engineering, the same question was settled in
favour of having \emph{both} unit and integration tests.

\textbf{``Decomposed evaluations are too costly.''} The dominant cost in current practice is the opportunity cost of misattributing improvements and selecting systems based on misleading signals. Even partial
decomposition is helpful: adding one component-level metric alongside the end-to-end score already improves the signal.

\textbf{``A reference solution is a reasonable
gold standard.''} It is, when the task is specified tightly enough
that the reference is the only reasonable shape. But the ambition
of agentic software engineering is broader than producing passing
patches: we want to evaluate design, abstraction choice, and
architectural fit --- qualities single reference patches do not encode and hidden test cannot see.

\section{Call to Action}
\label{sec:call}

We call on the community to act on the three remedies in
\S\ref{sec:symptoms}: report harness-aware metadata, move from
single-reference test sets to verifiers that admit multiple valid
solution shapes, and develop methods for component-level evaluation
alongside end-to-end scores. Each is tractable but non-trivial. And underneath
all three sits a harder problem: how do we state what we want a coding system to
do in terms an automated grader can apply, without prescribing how? This is the operationalisation gap of
\citet{wallach2025position} applied to agentic coding, and it is the
decisive constraint on the next generation of benchmarks. Until it
closes, benchmarks will keep grading agents on how closely they
resemble the solution that closed an issue --- whereas the ambition is
to let them do better.

\bibliographystyle{ACM-Reference-Format}
\bibliography{references}

@article{abstraction-adherence,
  title  = {A Proposed Evaluation Framework for Coding Agents: Tiles Enhance Proper Use of Public APIs by 35\%},
  author = {Shaposhnikov, Maksim and Gorinova, Maria I. and Willoughby, Rob and Knox, Dru},
  year   = {2025},
  publisher = {Tessl},
  journal = {Tessl Blog},
  urldate = {2025-11-12},
  note   = {\url{https://tessl.io/blog/proposed-evaluation-framework-for-coding-agents/}}
}

@article{evaluating-skills,
  title  = {A Proposed Framework For Evaluating Skills},
  author = {Shaposhnikov, Maksim},
  year   = {2025},
  publisher = {Tessl},
  journal = {Tessl Blog},
  urldate = {2026-04-15},
  note   = {\url{https://tessl.io/blog/a-proposed-framework-for-evaluating-skills-research-eng-blog/}}
}

@inproceedings{chen2021humaneval,
  title     = {Evaluating Large Language Models Trained on Code},
  author    = {Chen, Mark and Tworek, Jerry and Jun, Heewoo and Yuan, Qiming and Pinto, Henrique Ponde de Oliveira and Kaplan, Jared and Edwards, Harri and Burda, Yuri and Joseph, Nicholas and Brockman, Greg and others},
  year      = {2021},
  eprint    = {2107.03374},
  archivePrefix = {arXiv},
  primaryClass  = {cs.LG},
  note      = {\url{https://arxiv.org/abs/2107.03374}}
}

@article{austin2021mbpp,
  title   = {Program Synthesis with Large Language Models},
  author  = {Austin, Jacob and Odena, Augustus and Nye, Maxwell and Bosma, Maarten and Michalewski, Henryk and Dohan, David and Jiang, Ellen and Cai, Carrie and Terry, Michael and Le, Quoc and Sutton, Charles},
  year    = {2021},
  eprint  = {2108.07732},
  archivePrefix = {arXiv},
  primaryClass  = {cs.PL},
  note    = {\url{https://arxiv.org/abs/2108.07732}}
}

@inproceedings{jain2024livecodebench,
  title     = {{LiveCodeBench}: Holistic and Contamination-Free Evaluation of Large Language Models for Code},
  author    = {Jain, Naman and Han, King and Gu, Alex and Li, Wen-Ding and Yan, Fanjia and Zhang, Tianjun and Wang, Sida and Solar-Lezama, Armando and Sen, Koushik and Stoica, Ion},
  booktitle = {International Conference on Learning Representations (ICLR)},
  year      = {2025},
  eprint    = {2403.07974},
  archivePrefix = {arXiv},
  note      = {\url{https://arxiv.org/abs/2403.07974}}
}

@inproceedings{zhuo2024bigcodebench,
  title     = {{BigCodeBench}: Benchmarking Code Generation with Diverse Function Calls and Complex Instructions},
  author    = {Zhuo, Terry Yue and Vu, Minh Chien and Chim, Jenny and Hu, Han and Yu, Wenhao and Widyasari, Ratnadira and Yusuf, Imam Nur Bani and Zhan, Haolan and He, Junda and Paul, Indraneil and others},
  booktitle = {International Conference on Learning Representations (ICLR)},
  year      = {2025},
  eprint    = {2406.15877},
  archivePrefix = {arXiv},
  note      = {\url{https://arxiv.org/abs/2406.15877}}
}

@misc{many-swe-bench-passing-prs-would-not-be-merged-into-main,
    title = {Many SWE-bench-Passing PRs Would Not Be Merged into Main},
    author = {Parker Whitfill and Cheryl Wu and Joel Becker and Nate Rush},
    howpublished = {\url{https://metr.org/notes/2026-03-10-many-swe-bench-passing-prs-would-not-be-merged-into-main/}},
    year = {2026},
    month = {03},
}

@inproceedings{jimenez2024swebench,
  title     = {{SWE-bench}: Can Language Models Resolve Real-World {GitHub} Issues?},
  author    = {Jimenez, Carlos E. and Yang, John and Wettig, Alexander and Yao, Shunyu and Pei, Kexin and Press, Ofir and Narasimhan, Karthik},
  booktitle = {International Conference on Learning Representations (ICLR)},
  year      = {2024},
  eprint    = {2310.06770},
  archivePrefix = {arXiv},
  note      = {\url{https://arxiv.org/abs/2310.06770}}
}

@misc{openai2024swebenchverified,
  title  = {Introducing {SWE-bench Verified}},
  author = {{OpenAI}},
  year   = {2024},
  month  = {August},
  howpublished = {\url{https://openai.com/index/introducing-swe-bench-verified/}}
}

@misc{openai2026swebenchretire,
  title  = {Why {SWE-bench Verified} No Longer Measures Frontier Coding Capabilities},
  author = {{OpenAI}},
  year   = {2026},
  month  = {February},
  howpublished = {\url{https://openai.com/index/why-we-no-longer-evaluate-swe-bench-verified/}}
}

@inproceedings{yang2025swebenchmm,
  title     = {{SWE-bench Multimodal}: Do {AI} Systems Generalize to Visual Software Domains?},
  author    = {Yang, John and Jimenez, Carlos E. and Zhang, Alex L. and Lieret, Kilian and Yang, Joyce and Wu, Xindi and Press, Ori and Muennighoff, Niklas and Synnaeve, Gabriel and Narasimhan, Karthik R. and Yang, Diyi and Press, Ofir},
  booktitle = {International Conference on Learning Representations (ICLR)},
  year      = {2025},
  eprint    = {2410.03859},
  archivePrefix = {arXiv},
  note      = {\url{https://arxiv.org/abs/2410.03859}}
}

@misc{swebenchpro2025,
  title  = {{SWE-Bench Pro}: Can {AI} Agents Solve Long-Horizon Software Engineering Tasks?},
  author = {{Scale AI}},
  year   = {2025},
  eprint = {2509.16941},
  archivePrefix = {arXiv},
  note   = {\url{https://arxiv.org/abs/2509.16941}}
}

@article{aleithan2024swebenchplus,
  title  = {{SWE-Bench+}: Enhanced Coding Benchmark for {LLMs}},
  author = {Aleithan, Reem and Xue, Haoran and Mohajer, Mohammad Mahdi and Nnorom, Elijah and Uddin, Gias and Wang, Song},
  year   = {2024},
  eprint = {2410.06992},
  archivePrefix = {arXiv},
  note   = {\url{https://arxiv.org/abs/2410.06992}}
}

@article{wang2025solvedissues,
  title  = {Are ``Solved Issues'' in {SWE-bench} Really Solved Correctly? An Empirical Study},
  author = {Wang, You and Pradel, Michael and Liu, Zhongxin},
  year   = {2025},
  eprint = {2503.15223},
  archivePrefix = {arXiv},
  note   = {\url{https://arxiv.org/abs/2503.15223}}
}

@article{swebenchillusion2025,
  title  = {The {SWE-Bench} Illusion: When State-of-the-Art {LLMs} Remember Instead of Reason},
  author={Liang, Shanchao and Garg, Spandan and Moghaddam, Roshanak Zilouchian},
  year   = {2025},
  eprint = {2506.12286},
  journal={arXiv preprint arXiv:2506.12286},
  archivePrefix = {arXiv},
  note   = {\url{https://arxiv.org/abs/2506.12286}}
}

@inproceedings{liu2024agentbench,
  title     = {{AgentBench}: Evaluating {LLMs} as Agents},
  author    = {Liu, Xiao and Yu, Hao and Zhang, Hanchen and Xu, Yifan and Lei, Xuanyu and Lai, Hanyu and Gu, Yu and Ding, Hangliang and Men, Kaiwen and Yang, Kejuan and others},
  booktitle = {International Conference on Learning Representations (ICLR)},
  year      = {2024},
  eprint    = {2308.03688},
  archivePrefix = {arXiv},
  note      = {\url{https://arxiv.org/abs/2308.03688}}
}

@article{yao2024taubench,
  title  = {{$\tau$-bench}: A Benchmark for Tool-Agent-User Interaction in Real-World Domains},
  author = {Yao, Shunyu and Shinn, Noah and Razavi, Pedram and Narasimhan, Karthik},
  year   = {2024},
  eprint = {2406.12045},
  archivePrefix = {arXiv},
  note   = {\url{https://arxiv.org/abs/2406.12045}}
}

@article{chan2024mlebench,
  title  = {{MLE-bench}: Evaluating Machine Learning Agents on Machine Learning Engineering},
  author = {Chan, Jun Shern and Chowdhury, Neil and Jaffe, Oliver and Aung, James and Sherburn, Dane and Mays, Evan and Starace, Giulio and Liu, Kevin and Maksin, Leon and Patwardhan, Tejal and Madry, Aleksander and Weng, Lilian},
  year   = {2024},
  eprint = {2410.07095},
  archivePrefix = {arXiv},
  note   = {\url{https://arxiv.org/abs/2410.07095}}
}

@inproceedings{wijk2024rebench,
  title     = {{RE-Bench}: Evaluating Frontier {AI R\&D} Capabilities of Language Model Agents Against Human Experts},
  author    = {Wijk, Hjalmar and Lin, Tao and Becker, Joel and Jawhar, Sami and Parikh, Neev and Broadley, Thomas and Chan, Lawrence and Chen, Michael and Clymer, Joshua and Dhyani, Jai and others},
  booktitle = {International Conference on Machine Learning (ICML)},
  year      = {2025},
  eprint    = {2411.15114},
  archivePrefix = {arXiv},
  note      = {\url{https://arxiv.org/abs/2411.15114}}
}

@article{merrill2026terminalbench,
  title  = {{Terminal-Bench}: Benchmarking Agents on Hard, Realistic Tasks in Command Line Interfaces},
  author = {Merrill, Mike A. and Shaw, Alexander G. and Carlini, Nicholas and Li, Boxuan and others},
  year   = {2026},
  eprint = {2601.11868},
  archivePrefix = {arXiv},
  note   = {\url{https://arxiv.org/abs/2601.11868}}
}

@misc{frontierswe2026,
  title  = {{Frontier-SWE}: A Benchmark of Long-Horizon Software Engineering Tasks},
  author = {{Proximal Labs}},
  year   = {2026},
  howpublished = {\url{https://www.frontierswe.com/blog}}
}

@article{openai2025swelancer,
  title  = {{SWE-Lancer}: Can Frontier {LLMs} Earn \$1 Million from Real-World Freelance Software Engineering?},
  author = {{OpenAI}},
  year   = {2025},
  eprint = {2502.12115},
  archivePrefix = {arXiv},
  note   = {\url{https://arxiv.org/abs/2502.12115}}
}

@inproceedings{yang2024sweagent,
  title     = {{SWE-agent}: Agent-Computer Interfaces Enable Automated Software Engineering},
  author    = {Yang, John and Jimenez, Carlos E. and Wettig, Alexander and Lieret, Kilian and Yao, Shunyu and Narasimhan, Karthik and Press, Ofir},
  booktitle = {Advances in Neural Information Processing Systems (NeurIPS)},
  year      = {2024},
  eprint    = {2405.15793},
  archivePrefix = {arXiv},
  note      = {\url{https://arxiv.org/abs/2405.15793}}
}

@inproceedings{wang2025openhands,
  title     = {{OpenHands}: An Open Platform for {AI} Software Developers as Generalist Agents},
  author    = {Wang, Xingyao and Li, Boxuan and Song, Yufan and Xu, Frank F. and Tang, Xiangru and Zhuge, Mingchen and Pan, Jiayi and Song, Yueqi and Li, Bowen and Singh, Jaskirat and others},
  booktitle = {International Conference on Learning Representations (ICLR)},
  year      = {2025},
  eprint    = {2407.16741},
  archivePrefix = {arXiv},
  note      = {\url{https://arxiv.org/abs/2407.16741}}
}

@misc{aiderpolyglot,
  title  = {The Aider Polyglot Coding Benchmark},
  author = {Gauthier, Paul},
  year   = {2024},
  howpublished = {\url{https://aider.chat/2024/12/21/polyglot.html}}
}

@misc{codex,
  title        = {Introducing {Codex}},
  author       = {{OpenAI}},
  year         = {2025},
  howpublished = {\url{https://openai.com/index/introducing-codex/}}
}

@misc{claude-code,
  title        = {{Claude Code}},
  author       = {{Anthropic}},
  year         = {2025},
  howpublished = {\url{https://claude.com/product/claude-code}}
}

@misc{cursor-agent,
  title        = {{Cursor} Agents},
  author       = {{Cursor}},
  year         = {2025},
  howpublished = {\url{https://cursor.com/agents}}
}

@article{skillsbench2026,
  title  = {{SkillsBench}: Benchmarking How Well Agent Skills Work Across Diverse Tasks},
  author = {Li, Xiangyi and Chen, Wenbo and Liu, Yimin and others},
  year   = {2026},
  eprint = {2602.12670},
  archivePrefix = {arXiv},
  note   = {\url{https://arxiv.org/abs/2602.12670}}
}

@article{lee2026metaharness,
  title  = {{Meta-Harness}: End-to-End Optimization of Model Harnesses},
  author = {Lee, Yoonho and Nair, Roshen and Zhang, Qizheng and Lee, Kangwook and Khattab, Omar and Finn, Chelsea},
  year   = {2026},
  eprint = {2603.28052},
  archivePrefix = {arXiv},
  note   = {\url{https://arxiv.org/abs/2603.28052}}
}

@misc{openai2026harness,
  title  = {Harness Engineering},
  author = {{OpenAI}},
  year   = {2026},
  howpublished = {\url{https://openai.com/index/harness-engineering/}}
}

@misc{anthropic2025harness,
  title  = {Effective Harnesses for Long-Running Agents},
  author = {{Anthropic}},
  year   = {2025},
  month  = {November},
  howpublished = {\url{https://www.anthropic.com/engineering/effective-harnesses-for-long-running-agents}}
}

@inproceedings{wallach2025position,
  title     = {Position: Evaluating Generative {AI} Systems Is a Social Science Measurement Challenge},
  author    = {Wallach, Hanna and Desai, Meera and Cooper, A. Feder and Wang, Angelina and Atalla, Chad and Barocas, Solon and Blodgett, Su Lin and Chouldechova, Alexandra and Corvi, Emily and Dow, P. Alex and others},
  booktitle = {International Conference on Machine Learning (ICML)},
  year      = {2025},
  eprint    = {2502.00561},
  archivePrefix = {arXiv},
  note      = {\url{https://arxiv.org/abs/2502.00561}}
}

@inproceedings{jacobs2021measurement,
  title     = {Measurement and Fairness},
  author    = {Jacobs, Abigail Z. and Wallach, Hanna},
  booktitle = {Proceedings of the 2021 ACM Conference on Fairness, Accountability, and Transparency (FAccT)},
  year      = {2021},
  pages     = {375--385},
  doi       = {10.1145/3442188.3445901},
  note      = {\url{https://arxiv.org/abs/1912.05511}}
}

@misc{harnessvariance2026,
  title  = {Quantifying infrastructure noise in agentic coding evals},
  author = {Segato, Gian and {Engineering at Anthropic}},
  year   = {2026},
  howpublished = {\url{https://www.anthropic.com/engineering/infrastructure-noise}}
}

@misc{morphllm2025swepro,
  title  = {{SWE-Bench Pro}: A Detailed Analysis of Scaffold-Driven Score Variance},
  author = {{Morph Labs}},
  year   = {2025},
  howpublished = {\url{https://www.morphllm.com/swe-bench-pro}}
}

@misc{ai21scalingeval2025,
  title  = {Scaling Agentic Evaluation: Lessons from 200{,}000 {SWE-bench} Runs},
  author = {{AI21}},
  year   = {2025},
  howpublished = {\url{https://www.ai21.com/blog/scaling-agentic-evaluation-swe-bench/}}
}

@article{hassan2025agentic,
  title={Agentic software engineering: Foundational pillars and a research roadmap},
  author={Hassan, Ahmed E and Li, Hao and Lin, Dayi and Adams, Bram and Chen, Tse-Hsun and Kashiwa, Yutaro and Qiu, Dong},
  journal={arXiv preprint arXiv:2509.06216},
  year={2025}
}

@article{fan2025sweeffi,
  title  = {{SWE-Effi}: Re-Evaluating Software {AI} Agent System Effectiveness Under Resource Constraints},
  author = {Fan, Zhiyu and Vasilevski, Kirill and Lin, Dayi and Chen, Boyuan and Chen, Yihao and Zhong, Zhiqing and Zhang, Jie M. and He, Pinjia and Hassan, Ahmed E.},
  year   = {2025},
  eprint = {2509.09853},
  archivePrefix = {arXiv},
  note   = {\url{https://arxiv.org/abs/2509.09853}}
}

@article{li2025aidev,
  title  = {The Rise of {AI} Teammates in Software Engineering ({SE 3.0}): How Autonomous Coding Agents Are Reshaping Software Engineering},
  author = {Li, Hao and Zhang, Haoxiang and Hassan, Ahmed E.},
  year   = {2025},
  eprint = {2507.15003},
  archivePrefix = {arXiv},
  note   = {\url{https://arxiv.org/abs/2507.15003}}
}

@article{li2026aidev,
  title={{AIDev}: {S}tudying {AI} coding agents on {GitHub}},
  author={Li, Hao and Zhang, Haoxiang and Hassan, Ahmed E},
  journal={arXiv preprint arXiv:2602.09185},
  year={2026},
  note={\url{https://arxiv.org/abs/2602.09185}}
}

@article{gu2026peek,
  title  = {{PEEK}: Context Map as an Orientation Cache for Long-Context {LLM} Agents},
  author = {Gu, Zhuohan and Zhang, Qizheng and Khattab, Omar and Madden, Samuel},
  year   = {2026},
  eprint = {2605.19932},
  archivePrefix = {arXiv},
  note   = {\url{https://arxiv.org/abs/2605.19932}}
}

@article{gao2026decisionbench,
  title  = {{DecisionBench}: A Benchmark for Emergent Delegation in Long-Horizon Agentic Workflows},
  author = {Gao, Yuxuan and Wang, Megan and Yu, Yi Ling and Ma, Zijian Carl and Qu, Ao},
  year   = {2026},
  eprint = {2605.19099},
  archivePrefix = {arXiv},
  note   = {\url{https://arxiv.org/abs/2605.19099}}
}

@article{dehghani2021benchmarklottery,
  title={The benchmark lottery},
  author={Dehghani, Mostafa and Tay, Yi and Gritsenko, Alexey A and Zhao, Zhe and Houlsby, Neil and Diaz, Fernando and Metzler, Donald and Vinyals, Oriol},
  journal={arXiv preprint arXiv:2107.07002},
  year={2021}
}

@inproceedings{ribeiro2020checklist,
  title={Beyond accuracy: {B}ehavioral testing of {NLP} models with {CheckList}},
  author={Ribeiro, Marco Tulio and Wu, Tongshuang and Guestrin, Carlos and Singh, Sameer},
  booktitle={Proceedings of the 58th annual meeting of the association for computational linguistics},
  pages={4902--4912},
  year={2020}
}

@book{beck2002tdd,
  title={Test-driven development: by example},
  author={Beck, Kent},
  year={2002},
  publisher={Addison-Wesley Professional}
}

@misc{yang2026programbench,
      title={{ProgramBench}: Can Language Models Rebuild Programs From Scratch?}, 
      author={John Yang and Kilian Lieret and Jeffrey Ma and Parth Thakkar and Dmitrii Pedchenko and Sten Sootla and Emily McMilin and Pengcheng Yin and Rui Hou and Gabriel Synnaeve and Diyi Yang and Ofir Press},
      year={2026},
      eprint={2605.03546},
      archivePrefix={arXiv},
      primaryClass={cs.SE},
      url={https://arxiv.org/abs/2605.03546}, 
}

@misc{strongdm2025factory,
    title        = {{StrongDM} Software Factory},
    author       = {{StrongDM}},
    year         = {2025},
    howpublished = {\url{https://factory.strongdm.ai/}},
    note         = {Field notes on non-interactive agentic development}
}

@misc{scanlan2026intercom,
    title        = {How we use {Claude Code} today at {Intercom}},
    author       = {Brian Scanlan},
    year         = {2026},
    howpublished = {\url{https://www.linkedin.com/pulse/how-we-use-claude-code-today-intercom-brian-scanlan-eb7cc/}}
}

@book{sutton2018rlbook,
  title     = {Reinforcement Learning: An Introduction},
  author    = {Sutton, Richard S. and Barto, Andrew G.},
  edition   = {2nd},
  publisher = {MIT Press},
  year      = {2018}
}

@misc{symphony2026,
  title        = {An open-source spec for {Codex} orchestration: {Symphony}},
  author       = {Kotliarskyi, Alex and Zhu, Victor and Brock, Zach},
  year         = {2026},
  howpublished = {\url{https://openai.com/index/open-source-codex-orchestration-symphony/}}
}

@misc{gascity,
  title        = {{GasCity}},
  author       = {{Gastown Hall}},
  year         = {2026},
  howpublished = {\url{https://github.com/gastownhall/gascity}}
}

@misc{badertdinov2025swerebench,
      title={{SWE-rebench V2}: Language-Agnostic SWE Task Collection at Scale}, 
      author={Ibragim Badertdinov and Maksim Nekrashevich and Anton Shevtsov and Alexander Golubev},
      year={2026},
      eprint={2602.23866},
      archivePrefix={arXiv},
      primaryClass={cs.SE},
      note={\url{https://arxiv.org/abs/2602.23866}}
}

\end{document}